\documentclass[12pt]{article}
\pdfoutput=1
\usepackage{ifpdf}
\usepackage[a4paper,top=3.3cm,width=15.5cm, bottom=3.3cm, left=2.5cm]{geometry}

\usepackage[utf8x]{inputenc}
\usepackage{amssymb}
\usepackage{amsmath}
\usepackage{graphicx}
\usepackage{hyperref}
\usepackage{color}

\newcommand{\eq}{\begin{equation}}
\newcommand{\eqx}{\end{equation}}
\newcommand{\eqs}{\begin{equation*}}
\newcommand{\eqsx}{\end{equation*}}
\newcommand{\eqn}{\begin{eqnarray}}
\newcommand{\eqnx}{\end{eqnarray}}
\newcommand{\eqns}{\begin{eqnarray*}}
\newcommand{\eqnsx}{\end{eqnarray*}}

\newcommand{\bea}{\begin{eqnarray}}
\newcommand{\beal}[1]{\begin{eqnarray}\label{#1}}
\newcommand{\eea}{\end{eqnarray}}
\newcommand{\be}{\begin{equation}}
\newcommand{\bel}[1]{\begin{equation}\label{#1}}
\newcommand{\ee}{\end{equation}}
\newcommand{\rf}[1]{(\ref{#1})}
\newcommand{\dd}{ \textmd{d} }
\newcommand{\non}{\nonumber}

\newcommand{\f}[2]{\frac{#1}{#2}}

\newcommand{\om}{\omega}

\newcommand{\Dl}{\Delta}

\newcommand{\nn}{{\cal N}}

\usepackage[nodisplayskipstretch]{setspace}
\usepackage{authblk}

\title{{\bf Linearized nonequilibrium dynamics in nonconformal plasma}}
\author[1]{Romuald A. Janik\footnote{Email: {\tt romuald@th.if.uj.edu.pl}}}
\author[2]{Grzegorz Plewa\footnote{Email: {\tt g.plewa@ncbj.gov.pl}}}
\author[1]{Hesam Soltanpanahi\footnote{Email: {\tt hesam@th.if.uj.edu.pl}}}
\author[2,3]{Micha\l\ Spali\'nski\footnote{Email: {\tt mspal@fuw.edu.pl}}}
\affil[1]{Institute of Physics, Jagiellonian University, ul. {\L}ojasiewicza 11, 30-348 Krak\'ow, Poland}
\affil[2]{National Center for Nuclear Research, ul. Ho\.za 69, 00-681 Warsaw, Poland}
\affil[3]{Physics Department, University of Bia{\l}ystok, ul. Lipowa 41, 15-424 Bia{\l}ystok, Poland}
\date{}

\begin{document}

\maketitle
\thispagestyle{empty}

\begin{abstract}
We investigate the behaviour of the lowest
nonhydrodynamic modes in a class of holographic models which
exhibit an equation of state closely mimicking the one determined from 
lattice QCD. We compute the lowest quasinormal mode frequencies for a 
range of scalar self-interaction potentials and find that the damping of the
quasinormal modes at the phase transition/crossover falls off by a factor of around two from
conformality after factoring out standard conformal temperature dependence. The damping
encoded in the imaginary part of the frequencies
turns out to be correlated with the speed of sound and is basically
independent of the UV details of the model. We also find that the dynamics of the
nonhydrodynamic degrees of freedom remains ultralocal, even to a higher
degree, as we deviate from conformality. These results indicate that the
role of nonhydrodynamic degrees of freedom in the vicinity of the crossover transition
may be enhanced.
\end{abstract}

\newpage

\tableofcontents

\section{Introduction}

The quark-gluon plasma produced in relativistic heavy-ion collisions at RHIC and LHC
is very successfully described by phenomenological hydrodynamic models \cite{Shuryak:2014zxa}.
Nevertheless, it is quite clear that the hydrodynamic description is applicable
only from a certain initial time after an inherently nonequilibrium
initial phase of expansion. The estimation of this so-called `thermalization time'
has been at the focal point of much theoretical effort.

The main fundamental difficulty in studying such a problem from first
principles is that when the plasma is strongly coupled we need a method which
would be at the same time nonperturbative and which would work directly in
Minkowski signature.  For these reasons, the methods of the AdS/CFT
correspondence have been applied in this context although predominantly for
the case of the conformal $\nn=4$ SYM theory
\cite{Policastro:2001yc,Kovtun:2004de}.  Despite that, the results obtained
within this framework 
are very encouraging.

Firstly, the appearance of a hydrodynamical description of the resulting plasma
system is not an input but rather a result from a much more general dual
gravitational description which incorporates genuine nonhydrodynamical degrees
of freedom. Thus one can study the transition to hydrodynamics and its
properties.  Secondly, it has turned out that at the transition to
hydrodynamics, the components of the energy-momentum tensor in the local rest
frame are still significantly anisotropic, meaning that the plasma is then
still quite far from thermal equilibrium which indicates that the phrase
`early thermalization' used in this context is really a misnomer.  Thirdly,
for numerous initial conditions, the plasma behaves hydrodynamically with very
good accuracy when the dimensionless product of the proper time and
temperature $T\tau \sim 0.6-0.7$ \cite{Heller:2011ju, Jankowski:2014lna}.
Before that nonequilibrium degrees of freedom\footnote{In the present paper we
  use this term to denote all nonhydrodynamical degrees of freedom in the
  plasma.} are typically very important.

It is worth mentioning also numerous important results on the dynamics of
shock wave collisions \cite{Chesler:2010bi, Casalderrey-Solana:2013aba,
  Chesler:2015wra} which we do not describe in more detail here.

Since the full nonlinear dynamics in the deeply nonequilibrium regime is very
complicated as it is described by higher dimensional Einstein's equations and
can be studied essentially only using the methods of numerical general
relativity, it was suggested in \cite{Son:2002sd, Kovtun:2005ev} that it may
be useful to incorporate just the lowest, least damped nonhydrodynamic degrees
of freedom into the commonly used nonlinear hydrodynamic description.  On the
dual gravity side these degrees of freedom are the so-called quasinormal modes
(QNM) of a finite temperature black hole. The 4D equations involving these
nonequilibrium modes proposed in \cite{Heller:2014wfa} take as an input from
the gravitational description only their real and imaginary frequencies
$\om_{QNM}=\pm \om_R+i \om_I$.  Moreover, it turns out that the dependence of
these frequencies on the momentum $k$ is very mild and can be neglected in a
first approximation. This property leads to a certain `ultralocality' of the
dynamics of the nonequilibrium modes on top of a hydrodynamic flow.

All the above investigations were performed in the context of the conformal $\nn=4$ SYM
theory which, by its very definition does not exhibit any kind of phase transition or crossover behaviour.
It is thus very interesting to study what modifications to the above picture
appear for a nonconformal theory. There are various ways to model nonconformal theories
within the AdS/CFT correspondence either following a top-down approach by studying
a specific nonconformal theory with a known string theory construction, or
a bottom-up approach where the gravitational background is phenomenologically fixed
to give properties known from lattice QCD. In the present paper we decided to
concentrate on the latter approach partly for simplicity and partly in order to deal
with a gravitational system which has very similar equation of state to real QCD\footnote{A paper
investigating the complementary top-down approach appeared simultaneously~\cite{BHM}.}.

In this paper we will concentrate on the dynamics of the lowest nonhydrodynamic degrees of freedom
in the nonconformal setting. In particular we will investigate how the damping of these
modes changes when we approach the crossover temperature $T_c$ (defined more precisely later).
This answers an important question whether the role of nonhydrodynamic degrees of freedom
becomes more important or less important closer to the phase transition.
Secondly, we will investigate whether the `ultralocality' property observed for $\nn=4$ SYM
nonequilibrium degrees of freedom (QNM) still holds in the nonconformal case, especially close
to the crossover/phase transition.

The plan of this paper is as follows. In section 2 we will review the family of gravitational
backgrounds that we will consider. In sections 3 and 4 we will give some details on their explicit
numerical construction and on our procedure for finding quasinormal mode frequencies.
As far as we know such a procedure has not been employed so far in the literature and may
be useful also in other contexts. There we will also discuss the relation of scalar QNM with
the metric ones for these backgrounds and proceed, in section 5, to describe our results.
We close the paper with conclusions.

\section{The nonconformal gravity backgrounds}
\label{s.backgrounds}

In this paper we will study a family of black hole 
backgrounds which follow from an action of gravity coupled to a
single scalar field with a specific self interaction potential:
\eq
S=\int d^5x \sqrt{g}  \left[ R-\f{1}{2}\, \left( \partial \phi \right)^2 +V(\phi) \, \right]
\eqx
This family has been introduced by Gubser and collaborators in a series of
papers \cite{GUBSER, Gubser:2008yx, Gubser:2008sz} 
and used to mimic the QCD equation of state by
a judicious choice of the potential. These backgrounds
were then used to study bulk viscosity (which identically vanishes
in the conformal case) and quite recently in \cite{NORONHA}, where second order
hydrodynamic transport coefficients have been calculated. 
Ref. \cite{GUBSER} provided an \emph{approximate} but quite accurate formula
for computing the equation of state (or more precisely the speed of
sound~$c_s^2$), 
\eq
c_s^2= \f{d\log T}{d\log S}
\eqx
directly in terms of the scalar potential $V(\phi)$. Thus the scalar potential 
parametrizes the physics of the particular type of gauge theory plasma.

The family of scalar potentials that we consider is 
\eq
V(\phi)=  -12 \, \cosh (b\,\phi) +c_2 \, \phi^2+ c_4 \, \phi^4+ c_6 \, \phi^6\label{generalV}
\eqx
The quadratic terms in $\phi$ (mass term) determine, according to the standard AdS/CFT
dictionary, the dimension $\Delta$ of the operator
$\Delta\,(\Delta-4)=m^2\,L^2$, where $L$ is the AdS radius which we fix it to
one. 

\begin{figure}
\centerline{\includegraphics[height=6cm]{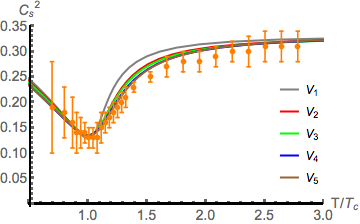}}
\caption{The speed of sound $c_s^2$ for the  potentials $V_1-V_5$ correspond to deformations of the theory by operators of dimensions 3.93, 3.67, 3.55, 3.10, 3.00 respectively given in table~1,  together with Lattice QCD data from \cite{Borsanyi:2012cr}.
\label{f.cs2}}
\end{figure}

In this paper we will mostly concentrate on a set of parameter choices
in~\rf{generalV} which approximately reproduce the equation of state of QCD plasma
as determined by 
the Budapest-Wuppertal group in \cite{Borsanyi:2012cr}. The resulting potentials are listed 
in table~\ref{t.pot} and the corresponding speed of sound is shown in
Fig.~\ref{f.cs2}. In that 
figure we show the speed of sound extracted from a numerical construction of
the corresponding black hole solution, as described in the following section,
together with the lattice QCD data for $c_s^2$. Note that in each case we are
free to choose the units of temperature. Here, following \cite{NORONHA} we fix this
freedom so that the
temperature corresponding to the lowest dip in $c_s^2$ coincides for all the
potentials. This will also be our provisional definition of the critical
temperature $T_c$. According to \cite{NORONHA} this value should be 143.8 MeV for
QCD. Note finally that for high temperatures, the equation of
state becomes essentially conformal. Similarly other properties such as the 
quasinormal frequencies also approach the conformal values characteristic
of $\nn=4$ SYM at high temperature.

In order to check that the qualitative conclusions are generic, we also considered
some other potentials leading to different profiles of $c_s^2(T)$. We will discuss them in section~\ref{s.other}.

\begin{table}
\centering
\begin{tabular}{c c c c c c}
\hline \hline
potential &  $ b $  &  $ c_2 $  &  $ c_4 $  &  $ c_6 $  &  $ \Delta $  \\
\hline
$V_1 $ & 0.606  &  2.06  & -0.1  & 0.0034 &  3.93  \\
$V_2 $ & 0.606  &  1.6  & -0.1  & 0.0034 &  3.67  \\
$V_3 $ & 0.606  &  1.4  & -0.1  & 0.0034 &  3.55  \\
$V_4 $ & 0.606  &  0.808  & -0.1  & 0.0034 &  3.10  \\
$V_5 $ & 0.606  &  0.703  & -0.1  & 0.0034 &  3.00  \\
\hline
\end{tabular}
\caption{The first group of potentials mimicking the QCD equation of state, together with the dimensions
$\Dl$ of the relevant scalar operators. Potential $V_6$ in table~2 was first constructed in \cite{GUBSER}, while
$V_{5}$ was recently used in \cite{NORONHA}.\label{t.pot}}
\end{table}

\section{Eddington-Finkelstein coordinates}

This section describes the black hole background solutions for the quasinormal
mode calculations. These backgrounds are the same as those in \cite{GUBSER},
but since our goal is to determine the quasinormal mode frequencies, it will 
be convenient to express them in Eddington-Finkelstein coordinates, rather
than in the coordinates used in \cite{GUBSER}. We will discuss this in more detail
in the following section on quasinormal modes.

The Ansatz for these solutions 
follows from the assumed symmetries: translation invariance in the
Minkowski directions as well as $SO(3)$ rotation symmetry in the spatial part. 
This leads to the following form of the line element:
\be
\label{linel0}
\dd s^2 = g_{tt} \dd t^2 + g_{xx} \dd \vec{x}^2 + g_{rr} 
\dd r^2 + 2 g_{rt} dr dt
\ee
where all the metric coefficients appearing in~\rf{linel0} are functions of the
radial coordinate $r$ 
alone, as is the scalar field $\phi$. This form of the field Ansatz
(determined so far only by the assumed symmetries) 
allows two gauge 
choices to be made. In 
\cite{GUBSER} a Schwarzschild-like gauge was adopted by taking $g_{tr}=0$,
accompanied by the condition $\phi = r$. 
The latter condition on the scalar field lead to key simplifications which
were 
used to solve the field 
equations. 
The final form of the 
Ansatz in \cite{GUBSER} was thus 
\beal{linelg}
\dd s^2 &=& e^{2 A} (-h \dd t^2 + \dd \vec{x}^2) + 
\frac{e^{2 B}}{h}  \dd r^2 \\
\phi &=& r
\eea
where $A$, $B$, and $h$ are functions of $r$ (or, equivalently $\phi$). 

For the purpose of computing the quasinormal modes it is very convenient to
use a different gauge -- the Eddington-Finkelstein gauge $g_{rr}=0$. It is
typically convenient to also impose the gauge choice  $g_{tr}=1$, but for our
purposes it turns out to be very effective to use the remaining gauge
freedom to set $\phi = r$. Furthermore, if we label the metric components as 
\beal{linelef}
\dd s^2 &=&  e^{2 A} (-h \dd t^2 + \dd \vec{x}^2) - 2 e^{A+B} \dd t \dd r \\
\phi &=& r
\eea
then the field equations take the form 
\begin{align}
\label{eq1}
A''-A'B'+\frac{1}{6}=0
\\[0ex]
\label{eq2}
h''+(4 A'-B')h'=0
\\[0ex]
\label{eq3}
6 A' h' + h(24 A'^2-1)+2 e^{2 B} V=0
\\[0ex]
\label{eq4}
4 A'-B'+\frac{h'}{h}-\frac{e^{2 B}}{h}V'=0,
\end{align}
where the prime denotes a derivative with respect to $\phi$. 

With the assumed labelling of metric coefficients~\rf{linelg}, equations
\rf{eq1}-\rf{eq4} are identical to those of appearing in \cite{GUBSER}.  
and so they can be solved following the method
described there (see also \cite{NORONHA}). In the remainder of this section we
review this procedure for completeness.

We are interested in solutions possessing a horizon, which requires that the
function $h$ should have a zero at some $\phi=\phi_H$:
\bel{hzero}
h(\phi_H) = 0,
\ee
It is easy to see that the solutions of Eq.~\rf{eq1}--\rf{eq4} can be
expressed in terms of a single function  
$G(\phi) \equiv A'(\phi)$:  
\begin{align}
\label{Asol}
A(\phi) = A_H + \int_{\phi_H}^\phi \dd \tilde{\phi} \,  G(\tilde{ \phi }),
\\[0ex]
\label{Bsol}
B(\phi) = B_H + \ln \left( \frac{G(\phi)}{G(\phi_H)} \right) +
\int_{\phi_H}^{\phi} \frac{\dd \tilde{\phi} }{6 G(\tilde{\phi})  }, 
\\[0ex]
\label{hsol}
h(\phi) = h_H+h_1 \int_{\phi_H}^{\phi} \dd \tilde{\phi} e^{-4 A(\tilde{\phi})+ B(\tilde{\phi})}
\end{align}
In the expressions above $A_H$,
$B_H$, $h_H$ and $h_1$ denote constants of integration which will be determined
by requiring the appropriate near-boundary behaviour and 
eq.~\rf{hzero}. 

As in \cite{GUBSER} by manipulating the field equations
\rf{eq1}-\rf{eq4} one 
finds the nonlinear ``master equation'' 
\bel{master}
\frac{G'}{G + V/3 V'} = \frac{\dd}{\dd \phi} \ln \left(
\frac{G'}{G}+\frac{1}{6 G}-4 G - \frac{G'}{G + V/3 V'}  \right) 
\ee
The strategy is to solve this equation numerically by integrating it from the
horizon at $\phi=\phi_H$ down toward the boundary at $\phi=0$. Once $G$ is
known, the metric coefficients can be recovered from Eq.~\rf{Asol}-\rf{hsol}.

Solving Eq.~\rf{master} requires appropriate boundary
conditions, which can be determined by evaluating \rf{eq3} and
\rf{eq4} at the horizon and using \rf{hzero}. In this way one finds
\be
V(\phi_H) = -3 e^{-2 B(\phi_H)} G(\phi_H) h'(\phi_H), \quad V'(\phi) = e^{-2
  B(\phi_H)} h'(\phi_H). 
\ee
From this it follows that
\be
\label{GH}
G(\phi_H) = -\frac{V(\phi_H)}{3 V'(\phi_H)}, 
\ee
Using~\rf{GH} and~\rf{master} one finds the following near-horizon expansion: 
\be
\label{Gexpo}
G(\phi) = -\frac{V(\phi_H)}{3 V'(\phi_H)} + \frac{1}{6} \left( \frac{V(\phi_H)
  V''(\phi_H)}{V'(\phi_H)^2}-1  \right)(\phi-\phi_H)+{\cal{O}}(\phi -
\phi_H)^2 
\ee
In particular, the expansion \rf{Gexpo} implies that 
\be
\label{GHprim}
G'(\phi_H) = \frac{1}{6} \left( \frac{V(\phi_H) V''(\phi_H)}{V'(\phi_H)^2}-1
\right). 
\ee
To summarize: to find a numerical solution of \rf{master} we can specify 
a value for $\phi_H$ and then use the conditions \rf{GH} and  \rf{GHprim} 
as boundary conditions for integrating \rf{master}. There is however one
technical  
complication in performing the numerical integration outlined above:
Eq. \rf{GH} implies that at the horizon  
\be
G(\phi_H)+V(\phi_H)/(3 V'(\phi_H))=0
\ee
which makes some terms of \rf{master}
singular. Even though such superficially singular
terms cancel, their presence makes numerical computations
troublesome. In order to circumvent this difficulty, instead of $\phi_H$ one
can initialize the integration at a point just outside the horizon, at  $\phi
= \phi_H -\epsilon_H$, where $\epsilon_H \ll 1$. Then using \rf{GHprim} one can
calculate $G(\phi_H - \epsilon_H)$ and 
$G'(\phi_H - \epsilon_H)$ and then use these values as the  boundary
conditions. One also needs to regularize at the boundary ($\phi=0$) by
integrating down to a small, but finite value $\phi=\epsilon_B$.   

Having determined $G$, one can find the metric from
\rf{Asol}--\rf{hsol}. The constants of integration can be determined following
\cite{GUBSER}. The result is 
\beal{iconst}
A_H &=& \frac{\ln \phi_H}{\Delta -4} + \int_0^{\phi_H} \dd \phi \left[
  G(\phi)-\frac{1}{(\Delta -4)\phi}  \right] \non\\
B_H &=& \ln \left(- \frac{4 V(\phi_H)}{V(0) V'(\phi_H) L}  \right) +
\int_0^{\phi_H} \frac{\dd \phi}{6 G(\phi)} \non\\
h_H &=& 0\non\\
h_1 &=& \frac{1}{\int_{\phi_H}^0 \dd \phi e^{-4 A(\phi)+B(\phi)}  }.
\eea
This way the metric is determined for any given choice of $\phi_H$.

The Beckenstein-Hawking formula for entropy leads to the following expression
for the entropy density
\bel{entrodens}
 s = \f{2\pi}{\kappa_5^2} e^{3A_H}
\ee
and the standard argument requiring non-singularity of the Euclidean
continuation at the horizon gives
\bel{temp}
 T = \f{e^{A_H - B_H} |h'(r_H)|}{4\pi} 
\ee
These equations lead to the formula
\bel{csapprox}
  c_s^2 = \f{d\log T / d\phi_H}{d\log s / d\phi_H}
    \approx \f{1}{3} - \f{1}{2} \f{V'(\phi_H)^2}{V(\phi_H)^2} \,.
\ee
where the latter equality, proposed in \cite{GUBSER}, is only approximate but works surprisingly well.
In the present paper when determining the speed of sound $c_s^2$, we always use 
the exact formula and determine it from $d\log T/d\log s$ with temperature and entropy
extracted from the exact numerical solutions.

\section{Quasinormal modes}
\label{sec.qnm}

\subsection{Introductory remarks}
\label{s.metricqnm}

It is convenient and enlightening to formulate the problem of finding
quasinormal modes in terms of gauge invariant variables, which are
diffeomorphism invariant  linear 
combinations of the perturbations. This approach is well known in general
relativity, and has been adopted in the holographic context in
\cite{Kovtun:2005ev}, where the conformal case of $\nn = 4$ supersymmetric
Yang-Mills theory was considered. The generalisation to the non-conformal
cases was undertaken in \cite{Benincasa:2005iv}. In this section we briefly
summarise our findings in the context of the models under consideration in
this paper. 

Under an infinitesimal diffeomorphism transformation, the metric and the
scalar field fluctuations 
transform as the metric and the scalar field itself, i.e. 
\be
g_{\mu \nu}\rightarrow g_{\mu \nu}-\nabla_\mu\, \xi_\nu-\nabla_\nu\,
\xi_\mu\,,\hspace{20mm} \phi\rightarrow \phi-\xi_\mu \nabla^\mu\, \phi. 
\ee
By examining linear combinations of the linearized perturbations one finds
five gauge invariant channels: two 
shear channels, one scalar channel, one sound channel and one bulk
channel\footnote{The bulk channel is a linear combination of transverse metric
  fluctuations and massive scalar fluctuation. One reason we call it "bulk"
  channel is it leads to non-zero bulk viscosity as well
  \cite{Gubser:2008yx}. One could refer to it as the "non-conformal" channel
  and to the rest,
  which are already known in conformal case, as conformal channels.}
\cite{Benincasa:2005iv}.

We assume the plane wave $e^{ -i \omega\, t + i \, k \, x }$ 
dependence of the fluctuations on boundary coordinates $t, x$ and some 
nontrivial dependence on the radial coordinate $r$. In general, the equations
for the shear and scalar channels are decoupled from the rest, which have the
same form as the conformal case. However the sound and bulk channels have
coupled second order equations \cite{Benincasa:2005iv}.

In the zero momentum limit, $k \rightarrow 0$, the equation for the sound mode
becomes decoupled from the bulk mode and at the same time the equations for
the other channels reduce to the equation for the QNM's of a massless
scalar field, except for the equation of the bulk channel which is still coupled
to the sound mode. Interestingly, this has two advantages. At $k=0$ it is enough
to find the QNM's for an external massless scalar field for the conformal
channels. On top of that, the coupling of the bulk mode with the sound mode means
that the former has the same frequency as the latter. So in this limit, all
the information about the QNM's of the theory is summarized in the QNM's of an
external massless scalar field. 

In view of this, the following analysis is focused on the QNM's of an external
massless scalar field in the nonconformal backgrounds under consideration. 
Of course, in general it would still be interesting to study the
metric and massive scalar field perturbations in detail.

\subsection{QNM of a massless scalar field}

In view of the results described in the previous section, we turn to exploring
the effects of conformal symmetry breaking by considering the QNM of a
massless scalar field $\Psi$ in the background \rf{linelef}. As discussed earlier,
the equation obtained for this case contains all the essential elements for
QNM perturbations of the background.

The field equation for a massless scalar is simply the wave equation 
\bel{eq.wave}
\nabla_A \nabla^A \Psi = 0
\ee
Quasinormal modes are solutions of the form 
\bel{qnmansatz}
\Psi = e^{- i \omega t+ i k x} \psi( \phi )
\ee
which satisfy the ingoing condition at the horizon, which in the
Eddington-Finkelstein coordinate system reduces to regularity there. 
Substituting~\rf{qnmansatz} into Eq.~\rf{eq.wave} and using the form of the
background metric given in Eq.~\rf{linelef} leads to the following
equation for the amplitude $\psi(\phi)$: 
\bel{eq.qnm}
(3 G \, V'+ V ) \psi'' -3 e^{-A-B} G' \left(e^{A+B} V'+2 i \omega \right) \psi' +
3 e^{-2 A-B} G' \left(k^2 e^{B}-3 i \omega  e^{A} G \right) \psi  = 0.
\ee
where primes denote
derivatives with respect to $\phi$. Note that only the functions $A$ and $B$
appear here -- the function $h$ drops 
out. Since we are imposing two boundary conditions, this is an eigenvalue
problem which can be solved only for specific values of the complex QNM
frequency $\omega$. 

\subsection{Numerical approach}
\label{sec.qnm.numer}

Quite generally, the chief advantage of using the Eddington-Finkelstein
coordinate system for finding 
quasinormal mode frequencies is twofold. Firstly, the ingoing boundary condition at the horizon
gets translated just to ordinary regularity of the solution at the horizon.
Secondly, due to the special form of the temporal part of Eddington-Finkelstein
metric, the dependence on the mode frequency of the relevant differential
equation 
is \emph{linear}. Hence the problem of finding the quasinormal frequencies amounts
to solving a linear ODE of the form
\bel{eq.gen.eig}
\hat{L}_1 \Psi = \om \hat{L}_2 \Psi
\ee
where $\hat{L}_1$ and $\hat{L}_2$ are specific differential operators,
with $\Psi$ satisfying essentially Dirichlet boundary conditions at the boundary
and being regular at the horizon. While many approaches to the problem of
finding quasinormal modes have been described in the literature \cite{Berti:2009kk}, we
believe that the approach we describe here is very effective in conjunction
with the spectral representation in terms of Chebyshev polynomials
\cite{Grandclement:2007sb}. This representation reduces the task of solving
Eq.~\rf{eq.gen.eig} to a set of linear equations. The differential operators
appearing in~\rf{eq.gen.eig} are represented as matrices, and due
to the linear dependence of this equation on $\omega$ this reduces to a
generalized matrix eigenvalue problem which can be solved very efficiently. 

In the case we are studying, the relevant equation is Eq.~\rf{eq.qnm}. 
This equation is indeed linear in $\omega$. We have 
implemented the strategy outlined in the previous paragraph and verified the
stability of the 
resulting solutions when varying the number of grid points in the Chebyshev
discretization. 
The 
results of these numerical calculations are presented in section~\ref{s.results}.

\section{Results}
\label{s.results}

We are interested in the dependence of the QNM frequencies on temperature and
$k$ in the vicinity of $k=0$. Therefore, in line with the discussion
in section~\ref{s.metricqnm}, we focus on the quasinormal modes of an external
massless scalar field. By use of the term ``external'' we wish to emphasize
that this scalar is \emph{distinct} from the scalar $\phi$ appearing in the
background geometry, as QNM modes of the latter are mixed with the metric
perturbations\footnote{See the discussion in section~\ref{s.metricqnm}}.

\subsection{The imaginary part of the QNM frequencies --- damping}

\begin{figure}
\centerline{\includegraphics[width=7cm]{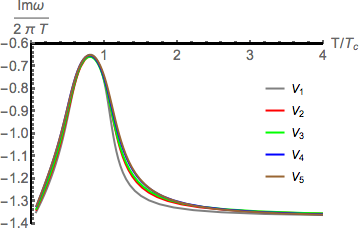}~~~~\includegraphics[width=7cm]{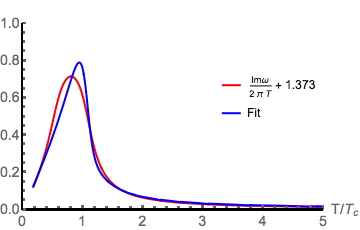}}
\caption{The imaginary parts of the lowest quasinormal mode at $k=0$ for the
  potentials from table~1 (left).  The imaginary part for potential $V_2$
  together with the ``phenomenological'' according to Eq.~\rf{e.fit}
  (right). 
\label{f.damp1}}
\end{figure}

In the left panel of figure~\ref{f.damp1} we show the imaginary parts of the
QNM frequencies in units of temperature 
which is the natural scale in the problem i.e.
\eq
\f{\text{Im}\, \om}{2 \, \pi \,  T}
\eqx
We observe that the damping significantly decreases (by a factor of 2) close to the transition.
This shows that in the
nonconformal case nonequilibrium dynamics become more important close to $T_c$. 
Moreover we find that the plots basically lie on top of each other for the
various potentials 
from table~1. This indicates that the QNM frequencies are not sensitive to the
fine details of the potentials 
but are essentially dependent just on the equation of state (speed of sound
$c_s^2(T)$), which was the common denominator of all the potentials from table~1.

In order to parameterize the dependence of the damping on deviation from
conformality, we propose a phenomenological formula expressing this 
as a linear combination of $c_s^2-\f{1}{3}$ and $T
\f{d}{dT}c_s^2(T)$. Specifically, we posit 
\eq
\label{e.fit}
\f{\text{Im}\, \om - \text{Im}\, \om_{\text{conf}}}{2\pi T}  = \gamma\, \left(c_s^2(T) - \f{1}{3}  \right) +
\gamma' \, T \f{d}{dT} c_s^2(T) 
\eqx
where $\gamma, \gamma'$ are phenomenological parameters and $\frac{\text{Im}\, \om_{\text{conf}}}{2 \, \pi \, T} =
-1.373$ is  the conformal limit value. These  
parameters can be fitted to the numerically calculated 
difference of the damping w.r.t the conformal case. 
For the potential $V_2$ in table~1 we got
\bel{eq.parvals}
\gamma =  -3.729 \, , \quad \gamma' =  0.452
\ee
In figure~\ref{f.damp1}(right) we show a plot of $(\text{Im}\, \om - \text{Im}\, \om_{\text{conf}})/(2\pi T) $ together with the fit.
This two parameter fit is surprisingly good and may be thus used phenomenologically to estimate
the damping in a nonconformal theory with the QCD equation of state. 
Since the quasinormal frequencies for the family of potentials we used in the
left panel of figure~\ref{f.damp1} basically coincide, the single choice of
parameters given in Eq.~\rf{eq.parvals} works well for all of them.

\subsection{The real part of the QNM frequencies}

\begin{figure}
\centerline{\includegraphics[width=7cm]{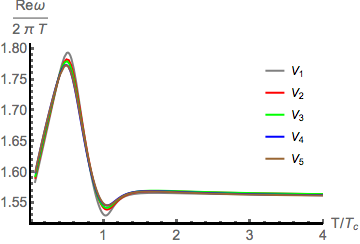}~~~~\includegraphics[width=7cm]{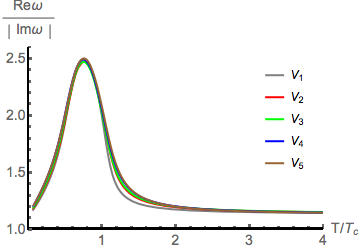}}
\caption{The real part of the lowest quasinormal mode at $k=0$ for the potentials from table~1 (left)
and the ratio of the real to the imaginary part (right).\label{f.real}}
\end{figure}

It is worth noting that similarly as for the imaginary part of the quasinormal
frequencies, the real parts of the frequencies corresponding to the various
potentials from table~1 are also very close to each other (see
figure~\ref{f.real}(left)). This signifies that the QNM frequencies are
basically insensitive to differences in the UV (since the various
potentials correspond to different $\Delta$'s) and are governed by IR
physics i.e. essentially the equation of state.

In figure~\ref{f.real}(right), we also see an enhancement of the real part of
the frequencies slightly below $T_c$.

\subsection{Ultralocality}

\begin{figure}
\centerline{\includegraphics[width=9cm]{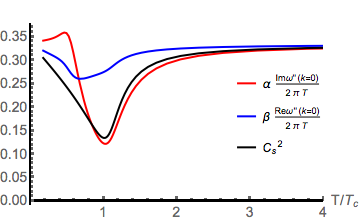}}
\caption{The curvature at $k=0$ of the damping frequency, overlaid with $\alpha \simeq \,1.114$ and $\beta\simeq 0.342$ for the  potential $V_2$ in table~1.\label{f.imomkk}}
\end{figure}

As indicated in the introduction, an interesting property of the dispersion relation for the
nonhydrodynamic degrees of freedom in the conformal case of $\nn=4$ SYM theory is the very mild
dependence of the frequencies $\om_I$ and $\om_R$ on the momentum $k$. In this section we show
that for nonconformal theories this property holds to an even higher degree. Interestingly,
the curvature of the damping i.e.
\eq
\f{\text{Im} \, \om''(k=0)}{2 \, \pi \, T}
\eqx 
follows to a surprising accuracy (at least until $T \sim T_c$, then it starts to deviate) 
just the speed of sound $c_s^2$, up to an overall numerical factor determined in the conformal
high temperature limit (i.e. equivalently in $\nn=4$ SYM).
The relevant plot is shown in figure~\ref{f.imomkk}. The plot for the other
potentials in the same family are basically the same.

\subsection{Further examples}
\label{s.other}

As discussed in section~\ref{s.backgrounds}, the choice of scalar potential translates
into a specific equation of state for the QCD-like gauge theory. In this
subsection we argue how the results discussed above apply to some additional
cases listed in table~\ref{t.pot.other}. 

\begin{table}
\centering
\begin{tabular}{c c c c c c}
\hline \hline
potential &  $ b $  &  $ c_2 $  &  $ c_4 $  &  $ c_6 $  &  $ \Delta $  \\
\hline
$V_6 $ & 0.606  &  2.06  & 0  & 0 &  3.93  \\
$V_7 $ & 0.606  &  1.8  & 0  & 0 &  3.79  \\
$V_8 $ & 0.606  &  1.145  & 0  & 0 &  3.37  \\
$V_9 $ & 0.606  &  0.808  & 0  & 0 &  3.10  \\
$V_{10} $ & 0.606  &  0.703  & 0  & 0 &  3.00  \\
$V_{11} $ &  $ 1/{ \sqrt{ 2 } } $  &  1.942  & 0  & 0 &  3.37  \\
\hline
\end{tabular}
\caption{The second group of potentials considered in the present paper.\label{t.pot.other}}
\end{table}

The potentials $V_6$ and  $V_{11}$ have been introduced in
\cite{GUBSER}. The former was used to mimic the QCD equation of state
using holography while the latter have a second-order phase transition at
$T=T_c$. Fixing $b=0.606$ corresponds to $c_s^2=0.15$ in the infrared
\cite{GUBSER}.  The variety of potentials leads to a range of conformal
weights $3\leq \Delta \leq 3.93$ (table \ref{t.pot.other}).

\begin{figure}
\centerline{\includegraphics[width=7cm]{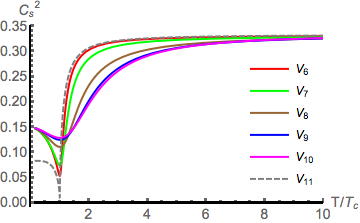}~~~~\includegraphics[width=7cm]{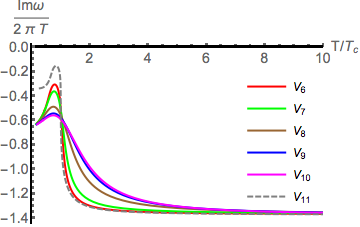}}
\caption{The speed of sound $c_s^2$ (left) and the imaginary parts of the lowest quasinormal modes at $k=0$ (right) for the potentials from table~2. The one with cusp in the $c_s^2$ (left) corresponds to the most decreased in damping (right). 
\label{other.cs2}}
\end{figure}

For all these cases the qualitative conclusions discussed earlier in this
section still hold. In figure~\ref{other.cs2} we plot the speed of sound
(left) and the imaginary parts of the lowest damped QNM's (right) for the
potentials $V_{6}-V_{11}$. Note that in all cases there is a critical
temperature, corresponds to the lowest value of $c_s^2$, which might be
considered as 
the cross over/phase transition point (related to the potentials). Damping of
the lowest QNM's decreases close to the $T=T_c$ by a factor of $2-7$
(depending on the potential) relative to the conformal theory at high
temperatures, figure \ref{other.cs2} (right). 

\begin{figure}
\centerline{\includegraphics[width=7cm]{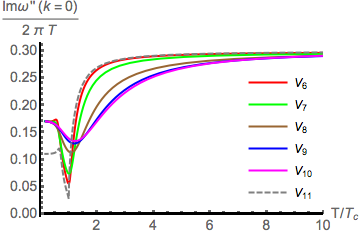}~~~~\includegraphics[width=7cm]{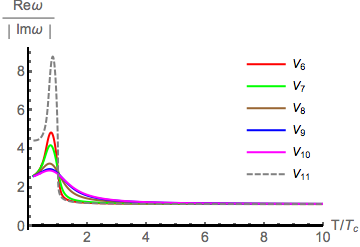}}
\caption{The curvature at $k=0$ for the damping frequency for potentials from
  table~2 (left) and the ratio of the real to the imaginary part (right) for
  the same potentials.}
\label{others2}
\end{figure}

Figure~\ref{others2} shows the momentum dependence of the imaginary 
part of QNM's (left) and the ratio of the real parts to the imaginary parts
(right). The former indicates that ultralocality still
holds for this more diverse class of potentials. Interestingly, comparing the
plots in figure \ref{other.cs2} (left) and \ref{others2} (left) suggest that
the phenomenological relation we found in previous section,\footnote{The
  coefficient $1.114$ is the same as "$\alpha$" introduced in figure
  \ref{f.imomkk}.} namely 
\be
c_s^2 \simeq 1.114\, \frac{\text{Im} \, \omega''(k=0)}{2 \, \pi \, T}
\ee
is valid for this class of potentials too (again untill $T\,\sim\, T_c$).

The ratio of real parts to the imaginary parts of the lowest QNM's in figure \ref{others2} (right) shows a decreasing at $T\, \sim \, T_c $ i.e. the lower $c_s^2$ at $T=T_c$ the bigger decreasing in $\text{Re}\, \om/\text{Im}\, \om$.

Surprisingly, even the potential $V_{11}$ with the second-order phase transition exhibits
qualitatively the same behaviour as the other potentials discussed in the present paper.

\section{Conclusions}

In this paper we carried out a study of the lowest quasi-normal modes in a
class of nonconformal holographic models which exhibit an equation of state
very similar to the one obtained using lattice QCD. This class of models was
introduced in \cite{GUBSER} and incorporates 5-dimensional gravity coupled to
a scalar field with a given self-interaction potential which parametrizes the
model.

The frequencies of the lowest quasinormal modes provide a scale for the
importance of nonhydrodynamic degrees of freedom, thus their determination is
of a definite phenomenological interest. Our main observations are the
following.

Firstly we found that, within the class of considered models, the imaginary
part of the QNM frequency is strongly correlated with the speed of sound
characteristic of the equation of state, once we factor out the trivial
conformal temperature dependence.  In particular, it decreases by a factor of
around two at the point of the QCD crossover transition.  This means that
nonequilibrium effects will become more pronounced closer to the QCD phase
transition/crossover.  This seems to be a robust characteristic of this class
of models and persists also for models with other equations of state
considered for completeness in section~\ref{s.other}. We provided a
phenomenological formula (\ref{e.fit}) linking the damping with the speed of
sound.  It is important to emphasize, that although the numerical values of
the coefficients in (\ref{e.fit}) are specific \emph{only} to the models
mimicking the QCD equation of state, similar fits, with coefficients of the
same sign and similar order of magnitude, work also for other models
considered in section~\ref{s.other}.

Secondly, we found that the quasinormal frequencies practically coincide for a
whole class of models (potentials) which lead to the same equation of state
(or more precisely to the same speed of sound $c_s^2(T)$ as a function of
temperature). In particular, they seem to be quite independent of the
particular UV properties of the concrete potential such as the anomalous
dimension of the operator deforming the theory.

Thirdly, we found that the property of ultralocality found in
\cite{Heller:2014wfa}, namely the very mild dependence of the quasinormal
modes on the momentum $k$ around $k=0$ persists away from conformality. Even
more so, it becomes more pronounced. We also noticed an intriguing feature
that the curvature of the imaginary part of the QNM frequencies around $k=0$
follows surprisingly well the speed of sound squared $c_s^2$.

We believe that the above observations should be of phenomenological interest, especially
as they indicate a more pronounced role of nonhydrodynamic degrees of freedom close
to the QCD phase transition/crossover. It would be very interesting to gain some
analytical understanding of these properties as well as to investigate directly
nonlinear dynamical evolution in such models.

\bigskip

{\bf Acknowledgements.} We thank Alex Buchel, Michał Heller and Rob Myers for
sharing a draft of their preprint prior to submission. HS and MS would like to
thank the organizers of the CERN-CKC TH Institute on Numerical Holography,
where a part of this research was carried out. RJ and HS wish to thank Galileo
Galilei Institute for Theoretical Physics for hospitality and the INFN for
partial support during the program \emph{Holographic Methods for Strongly
  Coupled Systems} where this work was finalized. RJ and HS were supported by
NCN grant 2012/06/A/ST2/00396. GP and MS were supported by NCN grant
2012/07/B/ST2/03794.

\end{document}